# Defending Saltwater Intrusion: The Freshwater Pushback

Xun Cai[1*], Qubin Qin[2*], Jian Shen[3], Holly A. Michael[4,5], Matthew L. Kirwan[3], Peter A. Raymond[1]

[1]School of the Environment, Yale University, New Haven, CT, 06511, USA.

[2]Coastal Studies Institute, East Carolina University, Wanchese, NC, 27981, USA.

[3]Virginia Institute of Marine Science, William & Mary, Gloucester Point, VA, 23062, USA

[4]Department of Earth Sciences, University of Delaware, Newark, DE, 19716, USA

[5]Department of Civil, Construction and Environmental Engineering, University of Delaware, Newark, DE, 19716, USA

[*]Contribute equally

Corresponding author: Xun Cai (xun.cai@yale.edu); ORCID: 0000-0002-4251-2384

**Abstract**

Saltwater intrusion (SWI) threatens freshwater availability, agriculture, and ecosystem resilience in coastal regions. While sea-level rise (SLR) is a known driver of long-term salinization, the counteracting role of freshwater discharge remains underexamined. Here, we combine long-term observations with numerical modeling and machine learning reconstruction to quantify the buffering capacity of freshwater outflows across the U.S. coastline. In systems such as Delaware Bay and parts of the Gulf and South Atlantic coasts, the salt front has shifted seaward in recent decades, linked to increased discharge, despite SLR over that time period. We show that a 10–35% increase in freshwater flow can offset the salinity impact of 0.5 m of SLR, though regional variation is significant. With future discharge trends diverging spatially, SWI responses will be highly uneven. These results highlight the critical role of freshwater management in mitigating salinity risks under climate change, with implications for water resource resilience, coastal planning, and long-term adaptation strategies.

**Scientific Significance Statement**

As sea levels rise (SLR) and climate pressures intensify, saltwater intrusion (SWI) has emerged as a critical issue for coastal communities, with implications for drinking water and food security, ecosystem integrity, and land use planning. This study provides a novel and quantitative assessment of the competing influences of freshwater discharge and SLR on SWI into surface water across diverse U.S. coastal systems. By integrating long-term observational records with numerical modeling and machine learning reconstruction, we challenge the prevailing assumption that SLR is the dominant long-term driver of SWI. Instead, this study reveals that changes in freshwater discharge can substantially buffer or even reverse the extent of salinization in many estuaries. These findings underscore the importance of hydrological variability in modulating coastal salinity patterns. Region-specific responses along the U.S. coastline – such as increased freshwater influence in the Gulf and South Atlantic versus intensifying SWI in the Mid-Atlantic and New England – highlight the need for tailored adaptation and management strategies. This work advances our mechanistic understanding of SWI under climate change and provides critical insight for water resource planning, ecosystem conservation, and coastal resilience efforts.

# 1 Introduction

Coastal zones – including estuaries, tidal rivers, and adjacent wetlands – are dynamic transition areas where freshwater and marine systems converge. These regions are characterized by substantial spatial and temporal variability in salinity, which plays an important role in shaping water quality. Elevated salinity levels in surface waters can threaten critical resources, including drinking water supplies and agricultural productivity (Barlow and Reichard, 2010; Li et al., 2025). Moreover, fluctuations in salinity directly affect ecosystem structure and function, influencing nutrient cycling, primary productivity, and the health and distribution of aquatic organisms (Yokoi et al., 2002; Berger et al., 2019; Kaushal et al., 2025). Among the key challenges facing these environments is saltwater intrusion (SWI), a process in which saline water encroaches into freshwater systems. Often associated with sea-level rise (SLR) and other factors such as drought, SWI is a critical coastal environment issue.

In estuarine systems, SWI is primarily controlled by the dynamic balance between oceanic saltwater inflow and upstream freshwater discharge – reflecting the interplay between baroclinic forcing and buoyancy-driven circulation (Hansen and Rattray, 1966; MacCready, 2004). Variability in salinity distributions and salt front positions (described by SWI length) is influenced by a suite of physical factors, including river inflow, estuarine geometry (length, depth), sea level, wind forcing, and human interventions (Wong and Valle-Levinson, 2002; MacCready and Geyer, 2010; Hong and Shen, 2012; Ralston and Geyer, 2019; Cook et al., 2023; Yang and Zhang, 2023; Liu et al., 2024; Wegman et al., 2024; Siemes et al., 2025). Among these factors, river inflow and SLR are considered as two key factors determing the SWI (Bellafiore et al., 2021; Lee et al., 2025). SLR can intensify SWI gradually by altering estuarine stratification and baroclinic pressure gradients over longer timescales (e.g., decades and centuries; Hilton et al., 2008; Najjar et al., 2010; Ross et al., 2015). Compared to SLR, freshwater discharge is considered to dominate short-term variability in SWI, particularly at seasonal to interannual timescales (MacCready and Geyer, 2010; Ralston and Geyer, 2019). Thus, numerous studies have employed empirical and analytical models based on discharge and gravitational circulation to estimate salt flux divergence, demonstrating reasonable skill in capturing estuarine salinity dynamics (Hansen and Rattray, 1965; Savenije, 1993; MacCready, 2004; Gay and O'Donnell, 2007).

While freshwater discharge is often considered a primary driver of SWI over short timescales, its influence can be equally, if not more, significant over longer periods. Sustained reductions in river flow – resulting from climate change, upstream water withdrawals, or land-use changes – can fundamentally alter estuarine salinity regimes and intensify SWI, even in the absence of notable SLR. For instance, studies have shown that decreasing river discharge poses a substantial climate-related risk to freshwater supplies, potentially surpassing the impacts of SLR in certain contexts (Lee et al., 2024 and 2025). Therefore, freshwater discharge should not be viewed solely as a short-term modulator of SWI but recognized as a critical and persistent driver whose long-term trends warrant equal attention in coastal vulnerability assessments and adaptation planning (Fig. 1).

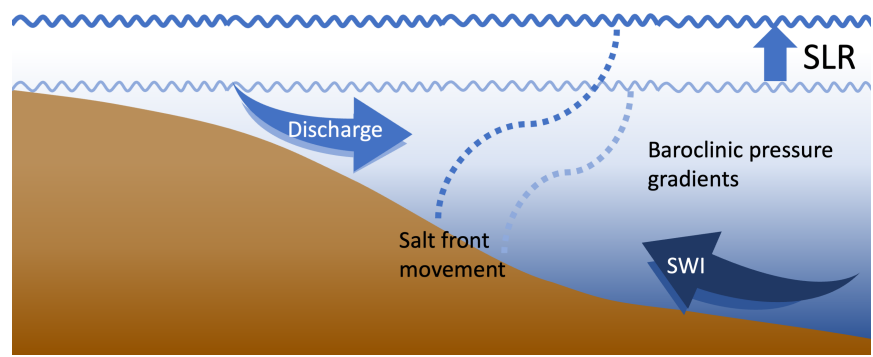

**Fig. 1**. Diagram illustrating the relative influence of freshwater discharge and SLR on surface water SWI in coastal regions.

Here we demonstrate the length of SWI ($L_s$) over long timescales to depend on river discharge ($Q$) and water depth ($H$), following a power-law relationship derived from both theoretical frameworks and observational studies (Monismith et al., 2002; MacCready and Geyer, 2010; Ralston and Geyer, 2019):

$$L_s \sim Q^{-\alpha} \cdot H^{\beta} \tag{1}$$

where $\alpha$ and $\beta$ are positive constants. This formulation indicates that increased freshwater discharge acts to shorten the intrusion length, while the greater water depth caused by SLR tends to extend it. Therefore, whether SWI becomes more or less extensive under future SLR scenarios depends on the relative magnitude of changes in freshwater discharge and estuarine depth.

The ratio of SWI length under any two conditions ($r_{L_s} = \frac{L_s'}{L_s}$) has the following expressions:

$$r_{L_s} \approx r_Q^{-\alpha} \cdot r_H^{\beta} \tag{2}$$

where $r_Q = \frac{Q'}{Q}$ and $r_H = \frac{H'}{H}$ are the corresponding ratios of changed river discharge and water depth, respectively.

Under SLR conditions (i.e., $r_H > 1$), the increase in water depth tends to promote further SWI. If this effect is fully counteracted by an increase in river discharge, the condition for a neutral SWI response becomes:

$$r_Q \approx r_H^{\frac{\beta}{\alpha}} \tag{3}$$

Several possible outcomes emerge depending on how freshwater discharge changes in response to SLR:

- If $r_Q \leq 1$, meaning freshwater discharge remains the same or decreases, then $r_{L_s} > 1$, indicating a landward migration of the salt front and an overall increase in SWI length.

- If $1 < r_Q < r_H^{\frac{\beta}{\alpha}}$, $r_{L_s} > 1$, freshwater discharge increases but not enough to fully counteract the deepening from SLR. In this case, $r_{L_s} > 1$, and SWI still extends farther inland, though the intrusion is partially mitigated.

- if $r_Q > r_H^{\frac{\beta}{\alpha}}$, $r_{L_s} < 1$, the increase in discharge exceeds the threshold needed to offset the effect of increased water depth. As a result, $r_{L_s} < 1$, and the salt front is pushed seaward, leading to a decrease in SWI length..

This analysis highlights that while enhanced freshwater discharge can mitigate the effects of SLR, it must increase sufficiently to counteract SWI. The effectiveness of this compensation varies across estuarine systems, governed by differences in morphology, tidal range, and mixing characteristics.

Based on this framework, we combine observational analysis, machine learning reconstruction, and numerical modeling experiments to investigate the role of freshwater discharge in modulating SWI under changing coastal conditions. We analyze decades of available observational data to assess spatial and temporal trends in estuarine salinity and salt front movements. Specifically, we reconstruct historical SWI length over the past century using a machine learning approach, allowing us to capture longer-term patterns and variability across U.S. coastal systems. Complementing this, we conduct scenario-based numerical experiments to explore the competing influences of SLR and freshwater discharge on SWI dynamics. Together, these methods demonstrate that sufficient freshwater discharge can offset – and in some cases reverse – the intensification of SWI driven by rising sea levels, highlighting the critical role of stream discharge management in future coastal resilience planning.

## 2 Methods

### 2.1 Available observations

We collected discharge records and specific conductivity data at surfaces water stations using datasets from the US Geological Survey (USGS; https://dashboard.waterdata.usgs.gov/app/nwd/en/). We applied the MATLAB toolbox (version 3_06_16) based on the Thermodynamic Equation of Seawater – 2010 (TEOS-10; https://github.com/TEOS-10/GSW-C) to calculate salinity in PSU from specific conductivity. In addition, we took the salinity observations from the National Estuarine Research Reserve System across the US (NERRS; https://cdmo.baruch.sc.edu/pwa/index.html). We obtained the regional Palmer Hydrological Drought Index (PHDI) from National Oceanic and Atmospheric Administration (NOAA) National Centers for Environmental information (NCEI; https://www.ncei.noaa.gov/access/monitoring/). The relative sea-level trends data is downloaded from NOAA Center for Operational Oceanographic Products and Services (CO-OPS, https://tidesandcurrents.noaa.gov/sltrends/sltrends.html).

**2.2 Numerical model and scenarios along the North American Atlantic Coast**

We employed 3D unstructured-grid NAAC (v1.0) model, which spans the Gulf of Maine, Mid-Atlantic Bight, and most of the South-Atlantic Bight along the North American Atlantic Coast, as the foundation for the SWI length reconstruction and numerical experiments (Cai et al., 2025). Built on the Semi-implicit Cross-scale Hydroscience Integrated System Model (SCHISM), NAAC (v1.0) provides evaluated 20-year simulations (2000–2020), capturing SLR signals and supporting subsequent machine learning applications.

To quantify the competing effects of SLR and freshwater discharge on SWI, we conducted 30 one-year numerical experiments using combinations of: (1) SLR levels of 0, 0.25, 0.5, 0.75, 1.0, and 1.5 m, and (2) freshwater discharge scaled to 0.8, 0.9, 1.0, 1.1, and 1.2 times the original volume. We selected the year 2008 as a baseline because it represents relatively average hydrological conditions – neither particularly wet nor dry. Each experiment included a one-year spin-up period to minimize the influence of initial conditions. Additionally, we ran supplemental scenarios with freshwater discharge at 0.8, 0.9, and 1.2 times the 2008 baseline, coupled with sea-level drops of 0.4 and 0.2 m. Notably, the scenario with 90% of 2008 discharge and a 0.4 m sea-level drop approximates projected conditions 100 years in the past and is used as a key reference case. The length of SWI in each estuary is estimated as the channel distance extending from the estuarine mouth to the location where bottom salinity falls to 1 PSU.

**2.3 Neural network model to re-build historical records**

The historical reconstruction of SWI length is based on a convolutional neural network (CNN) originally developed to forecast SWI length in the Chesapeake Bay (Shen et al., submitted). In this study, we applied the model in a hindcasting framework, validating its performance against available observational data dating back to the 1960s. The SWI length over time $Ls(t)$ is expressed as

$$Ls(t) = \widetilde{LS}(t,p) + Er(t) \tag{4}$$

where $\widetilde{LS}(t,p)$ is the simulations from SCHISM and $p$ is the model drivers (e.g., discharge, tide, surface water elevation, wind), and $Er(t)$ is the deviation between SCHISM simulation and $Ls(t)$. We take the empirical relationship based on MacCready and Geyer (2010):

$$\widetilde{Ls} \approx aQ^{-\gamma} \tag{5}$$

where $Q$ is discharge, $a$ is an adjusting coefficient, and $\gamma$ varies from 1 to 1/7. Therefore, SWI length can be written in the form of:

$$Ls(t) = \frac{a}{\Delta t}\int_{t-\Delta t}^{t} \widetilde{Ls}(\tau)\, d\tau + Er(t) \approx \frac{a}{\Delta t}\sum_{t-\Delta t}^{t} Q^{\gamma} + Er(t) \tag{6}$$

where $\Delta t$ is the backward integration interval representing accumulative effect of discharge. We trained the CNN model using SWI length simulations from the SCHISM model to estimate $Er(t)$. For each specific estuarine system, we identified the optimal number of days for calculating the cumulative backward mean flow that yields the strongest correlation with SWI length. We selected 16 $\Delta t$ that have highest correlation with $Ls$ backward three-month as input features to train the model. This approach has better predictive skill than simulate including past three-mounth daily flow when using CNN. Additionally, the scaling parameter $\gamma$ was found to be 1/5 for achieving the highest correlation at the Delaware Bay. Then we applied this trained model to reconstruct historical SWI length records using long-term USGS discharge data and sea-level signals at the estuary mouth. The hindcasted SWI length at Delaware Bay – especially the upstream extent during the historically severe intrusion events of the 1960s – was validated against observational records from the Delaware River Basin Commission (DRBC; https://www.nj.gov/drbc/), showing strong agreement and supporting the model's credibility.

**3 Observed divergence in SWI trends across U.S. coasts despite SLR**

In recent decades, coastal surface water salinity monitoring has expanded significantly across the U.S. shoreline. Among the 152 stations with salinity records since 1995 and extending at least 15 years, 62 stations (40.79%) show statistically significant trends ($p < 0.05$), while an additional 37 stations (24.34%) display possible trends ($p < 0.25$) (Table S1). Of the 99 stations with either significant or possible trends, 49 (49.49%) exhibit decreasing salinity, whereas 50 (50.51%) show increasing salinity (Table S1). The most pronounced annual decline in salinity, reaching –0.31 PSU per year, is observed at Pine Island, Florida. In contrast, Suisun Bay on the Pacific Coast displays the steepest increase at +0.27 PSU per year. These findings indicate that, despite rising sea levels over the past decades, a substantial number of U.S. estuaries are experiencing long-term reductions in salinity.

Focusing on the period from 2010 to the present, when continuous salinity monitoring has become more widespread, clearer regional patterns emerge (Fig. 2). Notably, decreasing salinity trends are concentrated in lower latitudes – particularly south of 37°N – and in specific hydrologic regions, especially along the South Atlantic and Gulf Coasts (Fig. 2). Among 64 stations across eight states in these regions, 58 show declining salinity: 33 with statistically significant trends, 15 with possible trends, and the remainder without significant changes. Conversely, on the Atlantic Coast north of 37°N, 38 of the 60 stations show increasing salinity, with 10 exhibiting significant trends and 9 with possible trends (Fig. 2). In the Mid-Atlantic region, divergent trends are observed: the main stems of the Chesapeake and Delaware Bays show increasing salinity, while their tributaries often display the opposite pattern. Regression analyses indicate that freshwater discharge within individual hydrologic zones or watersheds is a dominant driver of these regional salinity trends (Fig. 2b–f). Overall, the spatial distribution of salinity changes closely aligns with hydrologic boundaries, highlighting the strong influence of watershed-scale freshwater inputs on coastal salinity dynamics.

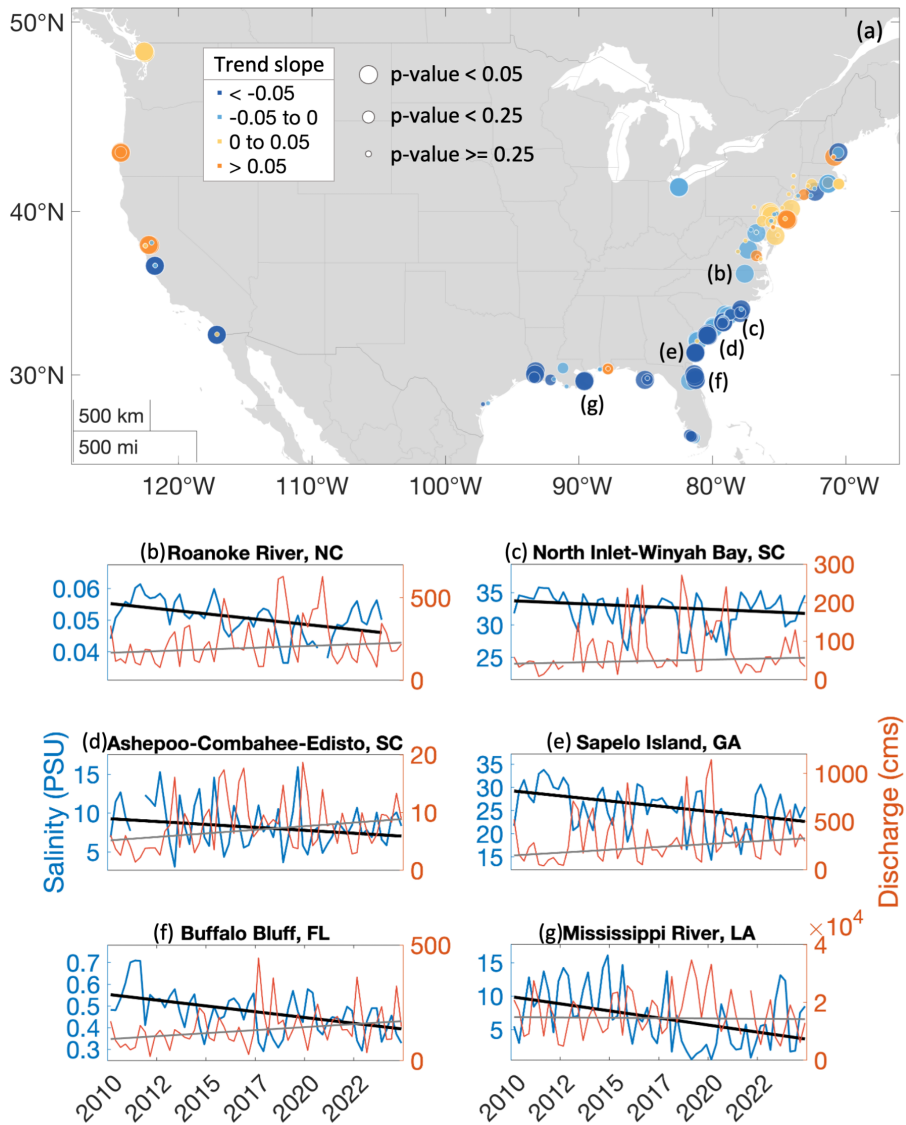

**Fig. 2**. Observed recent coastal annual salinity trends since 2010: (a) map with available stations showing significant/insignficant trends; (b-g) Time series of annual mean salinity and river discharge with trend lines at representative monitoring stations that show significant signals along the South Atlantic and Gulf Coasts.

Using reconstructed SWI length time series from Delaware Bay – one of the major estuaries along the U.S. East Coast, we identified a long-term decreasing trend in SWI length from the 1910s to 2024, with a linear slope of –0.036 km yr$^{-1}$ (Fig. 3a). This trend is statistically significant, with a p-value of 0.012. However, when the time series is segmented into shorter decadal periods, the direction and significance of the trends vary (Fig. S2). Notably, the elevated salinity levels and reduced freshwater input during the dry 1960s appear to dominate decadal variability, contributing to an overall increase in SWI length in the 1950s and a decrease of this trend in the 1970s.

Although SLR would be expected to increase SWI length over time in reference to the present-day coastline, the observed long-term trend runs counter to this expectation, with the SLR increasing slope of 0.004 m yr$^{-1}$ (Fig. 3d), suggesting the influence of other dominant factors. Indeed, we find a strong correlation between freshwater discharge whose trend slope is 0.36 cms yr$^{-1}$ (or the Palmer Hydrological Drought Index, PHDI, whose trend slope is 0.007 yr$^{-1}$) and SWI length (Fig. 3a-c). Across multiple decades, interannual and decadal variations in discharge explain much of the variability in SWI length. A first-difference regression analysis of the SWI length and discharge time series of annual mean level yields an $R^2$ of 0.79 with a p-value well below 0.005, demonstrating that even on century-long timescales, freshwater discharge is a dominant control on salt front dynamics at annual resolution.

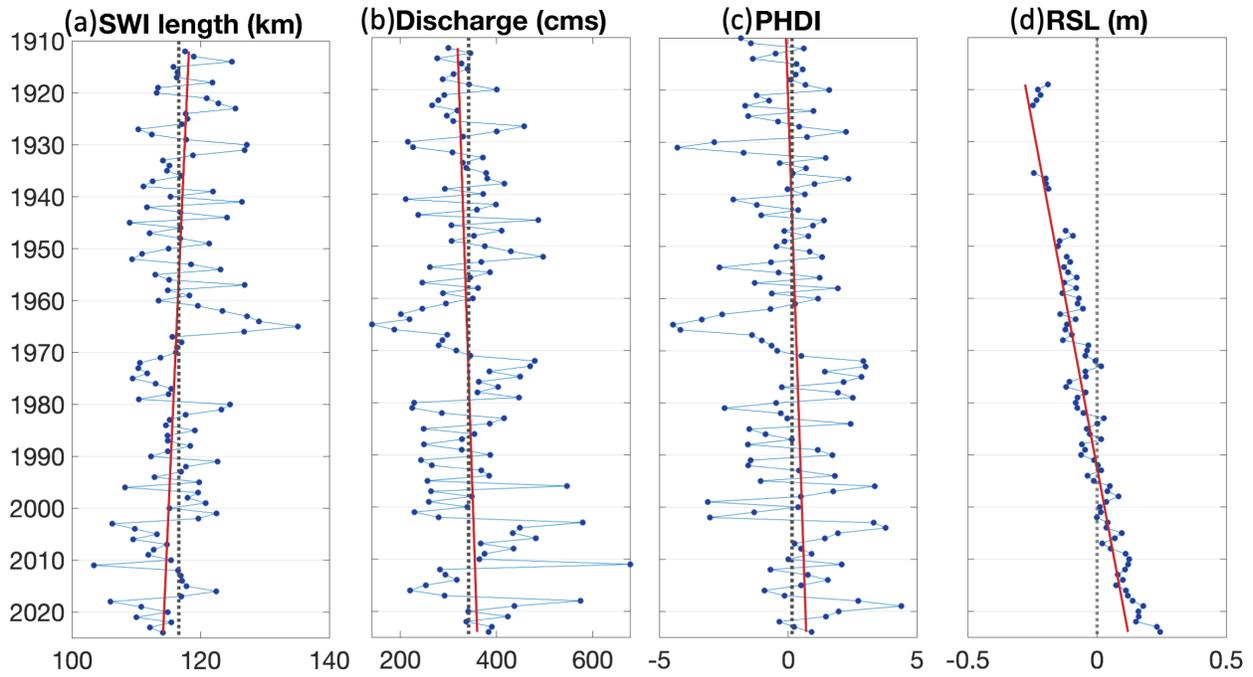

**Fig. 3**. Trends of (a) SWI length (km) to represent SWI, (b) discharge, (c) Palmer Hydrological Drought Index (PHDI), and (d) relative mean sea level to the most recent mean sea level at Delaware Bay in the past 110 years. There is a very strong and statistically significant negative linear relationship between (a) and (b) with a magnitude of 0.89.

## 4 Discussion

### 4.1 Competition between freshwater discharge and SLR

We utilized numerical experiments to further reveal this competitive interaction between SLR and freshwater discharge and estimate the corresponding $\alpha$ and $\beta$ at eq (1) in multiple coastal estuarine systems on the US East Coast (Table 1). Estuaries with shallower mean depths – such as the James River, Cape Fear River, and Winyah Bay – exhibit greater sensitivity to SLR. This heightened response arises from the relatively larger proportional increase in depth caused by a given amount of SLR, amplifying the effect on salt front dynamics.

Table 1. Estimated coefficients $\alpha$ and $\beta$ for estuaries in Fig. 4, p-values of each term are all smaller than $1 \times 10^{-6}$.

| Estuary | $\alpha$ | $\beta$ | $\beta/\alpha$ | Mean channel H (m) | r-square |
|---|---|---|---|---|---|
| Chesapeake Bay | 0.0507 | 0.4550 | 8.9770 | 20.1708 | 0.9951 |
| Delaware Bay | 0.1597 | 0.9153 | 5.7312 | 13.8485 | 0.9969 |
| James River | 0.2747 | 1.4909 | 5.4277 | 11.0749 | 0.9966 |
| Potomac River | 0.2160 | 1.2608 | 5.8368 | 14.1866 | 0.9901 |
| Cape Fear River | 0.4410 | 2.8203 | 6.3953 | 11.6521 | 0.9969 |
| Winyah Bay | 0.3166 | 1.0408 | 3.2877 | 7.0456 | 0.9967 |

Overall, a 0.5 m increase in sea level can be offset by increasing river discharge to approximately 10% to 35% of the original discharge volume, depending on estuarine characteristics (Fig. 4). Conversely, a 10% reduction in discharge can result in SWI changes equivalent to 0.2–0.5 m of SLR. Scenarios that combine both SLR and increased discharge generally show a moderated response in SWI due to this compensatory effect. Despite the overall pattern, the magnitude of change in SWI length varies across systems based on numerical modeling experiments (Fig. 4a). For example, in the Chesapeake Bay – a large estuary with an original SWI length of about 285 km – a 1 m rise in sea level results in a 6.4 km shift upriver, amounting to a 2.25% change (Fig. 4a). Delaware Bay, with a slightly shorter length (220–300 km) and a SWI length of 115 km, exhibits a comparable response, though it is more sensitive to barotropic influences due to its relatively well-mixed conditions (Fig. 4a). In general, larger estuaries tend to show more buffered responses to changes in SLR and discharge, while smaller systems exhibit more pronounced shifts in salinity structure, potentially driving ecosystem reorganization (Fig. 4a).

At the local scale, different salinity zones respond uniquely to the combined effects of SLR and discharge changes. In scenarios with the most extreme SWI – 1.5 m of SLR combined with an 80% reduction in discharge – mesohaline and oligohaline zones show salinity increases of 5.5 to 6.5 PSU (Fig. 4b). In contrast, changes in polyhaline zones (2 PSU) and tidal freshwater zones (1.8 PSU) are relatively modest. Most saltwater-dominated areas experience minimal change under these scenarios. Overall, the mesohaline and oligohaline regions are the most sensitive to the interplay between SLR and freshwater discharge, experiencing the largest shifts in local salinity.

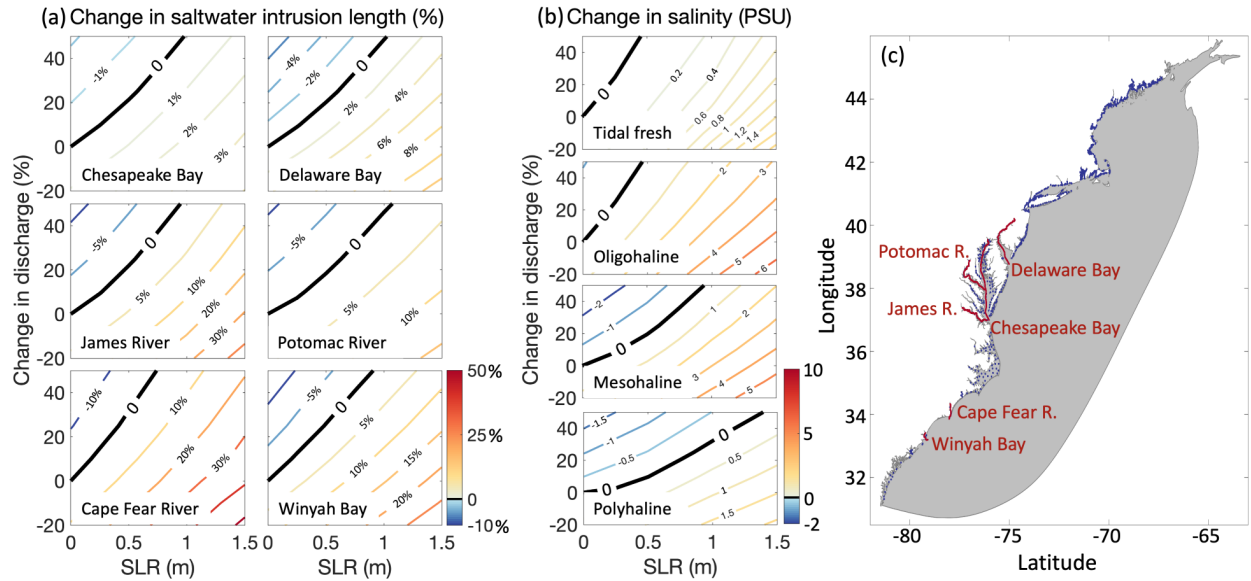

**Fig. 4**. (a) Simulated changes in salt front position under various SLR and river discharge scenarios at six estuaries along the US East Coast. (b) Simulated change in mean salinity in sampled shallow regions across different salinity zones, under varying combinations of SLR and discharge conditions. (c) Hydrodynamic model NAAC (v1.0) domain, with red lines indicating the estuarine transects corresponding to panels (a), and blue dots representing the sampling locations used in panel (b).

**4.2 Balancing climate impacts: how freshwater discharge offsets SWI across timescales**

Although SLR is widely recognized as a driver of long-term SWI (Hilton et a l., 2008; Najjar et al., 2010; Ross et al., 2015; O'Donnell et al., 2024), a significant number of observational records reveals a counterintuitive trend of freshening in many U.S. tidal rivers and estuaries over the recent decades (Fig. 2a). This freshening, particularly evident at the annual timescale, is primarily attributed to increasing freshwater discharge, which can offset the increase of salinity level typically associated with SLR (Fig. 2b-g).

Despite these long-term trends, short-term SWI events continue to occur across many of these "freshening" regions, especially during droughts and storm surge episodes. For example, in areas along the Gulf and South Atlantic Coasts, elevated salinity levels during recent drought events have surpassed the tolerance thresholds of salt-sensitive vegetation, resulting in ecological stress on marsh survival and altered biogeochemical cycling (Rahman et al., 2019; Herbert et al., 2015). These episodic events highlight the need to evaluate SWI using dynamic frameworks such as the salt wave concept (Cai et al., 2025), rather than relying solely on annual mean salinity levels. On shorter timescales (e.g., days to weeks), SWI is more strongly influenced by extreme weather events like storm surges and prolonged droughts.

Moreover, the spatial heterogeneity of SWI observed in this study, and corroborated by findings from other estuarine systems (e.g., Kaushal et al., 2025), indicates that increased freshwater discharge does not consistently buffer salinity across an entire estuary. In low-lying or subsiding regions, or areas more prone to marine inundation, high salinity zones may persist or expand despite regional discharge increases due to processes such as storm-surge overtopping (Frederiks et al., 2024). In addition to natural processes, upland water use can substantially impact river discharge (e.g., Peters et al., 2023), potentially offsetting climate-driven increases that mitigate SWI and highlighting the role of water management in combatting SWI.

In summary, while enhanced freshwater discharge offers mitigation of SWI under climate change, it does not fully eliminate the risks posed by episodic extremes or regional variability. An understanding of SWI dynamics across multiple timescales and spatial gradients is critical for anticipating ecosystem responses and managing coastal resilience.

**4.3 Potential future trends at global coasts**

While this study focuses on highlighting the role of freshwater discharge in mitigating or even reversing the effects of SLR on SWI, it does not explicitly simulate future SWI scenarios under projected climate change conditions. Our numerical experiments primarily explore the influence of discharge magnitude; however, future climate projections suggest that discharge variability may be shaped not only by changes in volume but also by altered timing and distribution – for example, longer drought intervals punctuated by more intense precipitation events. Thus the range of movement of the interfaces may be much greater in the future - creating potential issues temporally rather than on average. As a result, annual average discharge may not adequately capture the full range of SWI variability or its ecological consequences. This points to the need for future modeling efforts to incorporate more realistic hydrological patterns and climate-driven extremes to better understand how SWI may evolve and impact coastal ecosystems under changing environmental conditions.

Projections from the Community Earth System Model (CESM) suggest that future trends in river discharge will vary significantly across regions (Li et al., 2015; Lee et al., 2025). According to CESM-LE2 simulations, river discharge along the U.S. East Coast is expected to increase, which may help moderate or even weaken SWI in this region. In contrast, projections for other areas, such as the European Atlantic Coast, indicate a decline in river discharge, likely exacerbating SWI impacts there. These contrasting patterns suggest that coastal systems around the world may experience divergent responses to climate change, with some regions seeing reduced SWI risk and others facing intensified salinization challenges.

**4.4 Management Implications for Coastal Water Security**

Our results highlight that freshwater discharge is not only a short-term buffer against episodic saltwater intrusion but also a long-term determinant of coastal water security. Across the world, millions of people and extensive agricultural systems depend on freshwater derived from rivers, reservoirs, and aquifers hydraulically connected to estuaries. Studies have shown that reductions in river inflow can shift salinity fronts upstream, threatening municipal and irrigation intakes (e.g., Sacramento–San Joaquin Delta; Cloern & Jassby, 2012). Observational and modeling work also demonstrates that declining river discharge, together with sea-level rise, substantially increases intrusion risk in estuaries worldwide (Lee et al., 2025; Zamrsky et al., 2024).

Human interventions such as freshwater withdrawals already exert measurable impacts: while a single withdrawal project may elevate salinity only slightly, the cumulative effects of multiple withdrawals along the same river can significantly diminish the discharge reaching the coast (O'Donnell et al., 2024). For instance, in the James River (mean discharge ≈ 285 m³/s) at the East Coast of the U.S., model simulations show that a single proposed withdrawal of 40 MGD (~1.75 m³/s, ~0.6% of mean flow) could raise salinity in the oligohaline region by ~0.1 PSU, while three combined projects totaling 88 MGD (~3.84 m³/s, ~1.3% of mean flow) could produce salinity increases of ~0.2 PSU (Qin et al., 2025). Although the effect of one project may appear modest, the cumulative impacts demonstrate how even relatively small fractions of mean flow, when compounded, can substantially erode the natural freshwater pushback against saltwater intrusion. This compounding effect becomes particularly critical when considered alongside projected sea-level rise, which will further increase the hydraulic pressure driving saltwater intrusion (Moore & Joye, 2021).

Large dams and reservoirs, by redistributing freshwater seasonally, also influence estuarine salinity. In many cases, regulation reduces natural baseflow and mismatches between release schedules and salinity pressure weaken freshwater pushback (White and Kaplan, 2017). At the same time, strategically timed releases could enhance resilience if coordinated with estuarine thresholds. Other anthropogenic pressures, including coastal groundwater pumping (Werner et al., 2013), channel deepening (Ralston et al., 2019), land use change (Bhattachan et al., 2018), and inter-basin diversions (Yi & Kondolf, 2024; Cloern et al., 2017), further modify freshwater inflows and intrusion dynamics, underscoring the need for integrated, long-term planning.

These findings suggest that water management strategies must move beyond evaluating withdrawals and reservoir operations under current flow capacity alone. Instead, sustainable allocation will require incorporating long-term discharge trends, cumulative watershed interventions, and future sea-level rise into planning. This perspective is reinforced by recent synthesis studies highlighting the need to anticipate combined pressures of declining flows and sea-level rise, rather than assuming current allocation rules will remain viable (Lee et al., 2025; Missimer, 2025). Establishing discharge thresholds that account for both present and projected conditions would provide a forward-looking framework for safeguarding freshwater supplies. Anticipating these combined pressures is essential to design resilient policies that align with national priorities for sustainable coastal water security.

# 5 Conclusions

This study highlights the critical role of freshwater discharge in mitigating the effects of SLR on SWI across U.S. coastal regions. Despite an observed SLR of approximately 0.5 meters over the past century, many major estuaries along the U.S. East Coast have exhibited long-term decreasing trends in salt front distance. Over recent decades, SWI trends have varied considerably among coastal systems, reflecting differences in regional hydrology. Our findings demonstrate that freshwater discharge is not only a key driver of SWI variability on short-term (e.g., seasonal) scales but also exerts substantial influence over long-term trajectories. Looking ahead, the combined and potentially competing effects of changing freshwater discharge and continued SLR are likely to produce complex and regionally distinct SWI responses, emphasizing the need for integrated watershed–coastal management strategies.

## Acknowledgments

This research was supported by an NSF OCE-PRF fellowship (grant no. 2403359).

## Data availability

All the salinity data are publicly available from USGS (https://dashboard.waterdata.usgs.gov/app/nwd/en/), NERR (https://cdmo.baruch.sc.edu/pwa/index.html), NOAA CO-OPS (https://tidesandcurrents.noaa.gov/sltrends/sltrends.html), and NOAA NCEI (https://www.ncei.noaa.gov/access/monitoring/). Scripts for case study data analysis can be accessed at https://doi.org/10.5281/zenodo.17101625 (Cai et al., 2025).

## Author contributions

XC and QQ conceived the study and developed the overall concept. XC and QQ collected observational data and performed data analysis. JS and XC contributed to the historical data reconstruction. XC carried out the numerical model experiments. QQ and XC collaborated on equation development and result interpretation. HM, MK, and PR provided guidance on application and discussion. XC and QQ prepared the initial manuscript draft. All authors contributed to manuscript writing and revisions.

# References

Barlow, P.M. & Reichard, E.G. (2010). Saltwater intrusion in coastal regions of North Barlow, P.M. & Reichard, E.G. (2010). Saltwater intrusion in coastal regions of North America. *Hydrogeology Journal*, *18*(1), pp.247-260. doi: 10.1007/s10040-009-0514-3

Bellafiore, D., Ferrarin, C., Maicu, F., Manfè, G., Lorenzetti, G., Umgiesser, G., ... & Levinson, A. V. (2021). Saltwater intrusion in a Mediterranean delta under a changing climate. *Journal of Geophysical Research: Oceans*, 126(2), e2020JC016437. doi: 10.1029/2020JC016437

Berger, E., Frör, O., & Schäfer, R.B. (2019). Salinity impacts on river ecosystem processes: a critical mini-review. *Philosophical Transactions of the Royal Society B*, *374*(1764), p.20180010. doi: 10.1098/rstb.2018.0010

Bhattachan, A., Emanuel, R. E., Ardón, M., Bernhardt, E. S., Anderson, S. M., Stillwagon, M. G., ... & Wright, J. P. (2018). Evaluating the effects of land-use change and future climate change on vulnerability of coastal landscapes to saltwater intrusion. *Elem Sci Anth*, 6, 62. doi: 10.1525/elementa.316

Cai, X., Qin, Q., Kirwan, M., Michael, H., Shen, J., Mach, K. J., & Raymond, P. (2025). Recognizing Salt Wave Events in Coastal Systems. *arXiv preprint*. doi: 10.48550/arXiv.2505.07127

Cai, X., Qin, Q., Cui, L., Yang, X., Zhang, Y. J., & Shen, J. (2025). NAAC (v1. 0): A Seamless Two-Decade Cross-Scale Simulation from the North American Atlantic Coast to Tidal Wetlands Using the 3D Unstructured-grid Model SCHISM (v5. 11.0). *EGUsphere*, 2025, 1-23. doi: egusphere-2025-593

Cai, X., Qin, Q., Shen, J., Michael, H., Kirwan, M., and Raymond, P. (2025), Defending Saltwater Intrusion: The Freshwater Pushback. *Zenodo*. doi: 10.5281/zenodo.17101625

Cloern, J. E., Abreu, P. C., Carstensen, J., Chauvaud, L., Elmgren, R., Grall, J., ... & Yin, K. (2016). Human activities and climate variability drive fast-paced change across the world's estuarine–coastal ecosystems. *Global change biology*, 22(2), 513-529. doi: 10.1111/gcb.13546

Cloern, J. E., & Jassby, A. D. (2012). Drivers of change in estuarine-coastal ecosystems: Discoveries from four decades of study in San Francisco Bay. *Reviews of Geophysics*, 50(4). doi: 10.1029/2012RG000397

Cook, S. E., Warner, J. C., & Russell, K. L. (2023). A numerical investigation of the mechanisms controlling salt intrusion in the Delaware Bay estuary. *Estuarine, Coastal and Shelf Science*, 283, 108257. doi: 10.1016/j.ecss.2023.108257

Frederiks, R. S., Paldor, A., Carleton, G., & Michael, H. A. (2024). A comparison of sea-level rise and storm-surge overwash effects on groundwater salinity of a barrier island. *Journal of Hydrology*, 644, 132050. doi: 10.1016/j.jhydrol.2024.132050

Gay, P. S., & O'Donnell, J. (2007). A simple advection-dispersion model for the salt distribution in linearly tapered estuaries. Journal of Geophysical Research: Oceans, 112(C7). doi: 10.1029/2006JC003840

Hansen, D. V., & Rattray Jr, M. (1966). New dimensions in estuary classification 1. *Limnology and oceanography*, 11(3), 319-326. doi: 10.4319/lo.1966.11.3.0319

Herbert, E. R., Boon, P., Burgin, A. J., Neubauer, S. C., Franklin, R. B., Ardón, M., ... & Gell, P. (2015). A global perspective on wetland salinization: ecological consequences of a growing threat to freshwater wetlands. *Ecosphere*, 6(10), 1-43. doi: 10.1890/ES14-00534.1

Hilton, T. W., Najjar, R. G., Zhong, L., & Li, M. (2008). Is there a signal of sea-level rise in Chesapeake Bay salinity?. *Journal of Geophysical Research: Oceans*, 113(C9). doi: 10.1029/2007JC004247

Hong, B., & Shen, J. (2012). Responses of estuarine salinity and transport processes to potential future sea-level rise in the Chesapeake Bay. *Estuarine, coastal and shelf science*, 104, 33-45. doi: 10.1016/j.ecss.2012.03.014

Kaushal, S. S., Shelton, S. A., Mayer, P. M., Kellmayer, B., Utz, R. M., Reimer, J. E., ... & Chant, R. J. (2025). Freshwater faces a warmer and saltier future from headwaters to coasts: climate risks, saltwater intrusion, and biogeochemical chain reactions. *Biogeochemistry*, 168(2), 31. doi: 10.1007/s10533-025-01219-6

Lee, J., Biemond, B., de Swart, H., & Dijkstra, H. A. (2024). Increasing risks of extreme salt intrusion events across European estuaries in a warming climate. *Communications Earth & Environment*, 5(1), 60. doi: 10.1038/s43247-024-01225-w

Lee, J., Biemond, B., van Keulen, D., Huismans, Y., van Westen, R. M., de Swart, H. E., ... & Kranenburg, W. M. (2025). Global increases of salt intrusion in estuaries under future environmental conditions. *Nature Communications*, 16(1), 3444. doi: 10.1038/s41467-025-58783-6

Li, H. Y., Leung, L. R., Getirana, A., Huang, M., Wu, H., Xu, Y., ... & Voisin, N. (2015). Evaluating global streamflow simulations by a physically based routing model coupled with the community land model. *Journal of Hydrometeorology*, 16(2), 948-971. doi: 10.1175/JHM-D-14-0079.1

Li, M., Najjar, R. G., Kaushal, S., Mejia, A., Chant, R. J., Ralston, D. K., ... & Wang, X. (2025). The Emerging Global Threat of Salt Contamination of Water Supplies in Tidal Rivers. *Environmental Science & Technology Letters*. doi: 10.1021/acs.estlett.5c00505

Liu, J., Hetland, R., Yang, Z., Wang, T., & Sun, N. (2024). Response of salt intrusion in a tidal estuary to regional climatic forcing. *Environmental Research Letters*, 19(7), 074019. doi: 10.1088/1748-9326/ad4fa1

MacCready, P. (2004). Toward a unified theory of tidally-averaged estuarine salinity structure. *Estuaries*, 27, 561-570. doi: 10.1007/BF02907644

MacCready, P., & Geyer, W. R. (2010). Advances in estuarine physics. *Annual review of marine science*, 2(1), 35-58. doi: 10.1146/annurev-marine-120308-081015

Missimer, T. M., & Maliva, R. G. (2025). Salinity Barriers to Manage Saltwater Intrusion in Coastal Zone Aquifers During Global Climate Change: A Review and New Perspective. *Water*, 17(11), 1651. doi: 10.3390/w17111651

Monismith, S. G., Kimmerer, W., Burau, J. R., & Stacey, M. T. (2002). Structure and flow-induced variability of the subtidal salinity field in northern San Francisco Bay. *Journal of physical Oceanography*, 32(11), 3003-3019. doi: 10.1175/1520-0485(2002)032<3003:SAFIVO>2.0.CO;2

Moore, W. S., & Joye, S. B. (2021). Saltwater intrusion and submarine groundwater discharge: acceleration of biogeochemical reactions in changing coastal aquifers. *Frontiers in Earth Science*, 9, 600710. doi: 10.3389/feart.2021.600710

Najjar, R. G., Pyke, C. R., Adams, M. B., Breitburg, D., Hershner, C., Kemp, M., ... & Wood, R. (2010). Potential climate-change impacts on the Chesapeake Bay. *Estuarine, Coastal and Shelf Science*, 86(1), 1-20. doi: 10.1016/j.ecss.2009.09.026

O'Donnell, K. L., Bernhardt, E. S., Yang, X., Emanuel, R. E., Ardón, M., Lerdau, M. T., ... & Wright, J. P. (2024). Saltwater intrusion and sea level rise threatens US rural coastal landscapes and communities. *Anthropocene*, 45, 100427. doi: 10.1016/j.ancene.2024.100427

Peters, C. N., Kimsal, C., Frederiks, R. S., Paldor, A., McQuiggan, R., & Michael, H. A. (2022). Groundwater pumping causes salinization of coastal streams due to baseflow depletion: Analytical framework and application to Savannah River, GA. *Journal of Hydrology*, 604, 127238. doi: 10.1016/j.jhydrol.2021.127238

Qin, Q., Cai, X., and Shen, J. (2025). Cumulative Impacts of Multiple Municipal Water Withdrawals on Estuarine Salinity Dynamics: A Case Study in the James River.

Rahman, M. M., Penny, G., Mondal, M. S., Zaman, M. H., Kryston, A., Salehin, M., ... & Müller, M. F. (2019). Salinization in large river deltas: Drivers, impacts and socio-hydrological feedbacks. *Water security*, 6, 100024. doi: 10.1016/j.wasec.2019.100024

Ralston, D. K., & Geyer, W. R. (2019). Response to channel deepening of the salinity intrusion, estuarine circulation, and stratification in an urbanized estuary. *Journal of Geophysical Research: Oceans*, 124(7), 4784-4802. doi: 10.1029/2019JC015006

Ross, A.C., Najjar, R.G., Li, M., Mann, M.E., Ford, S.E., & Katz, B. (2015). Sea-level rise and other influences on decadal-scale salinity variability in a coastal plain estuary. *Estuarine, Coastal and Shelf Science*, 157, pp.79-92. doi: 0.1016/j.ecss.2015.01.022

Savenije, H. H. (1993). Predictive model for salt intrusion in estuaries. *Journal of Hydrology*, 148(1-4), 203-218. doi: 10.1016/0022-1694(93)90260-G

Shen, J., Cai, X., and Qin, Q. (submitted). A Machine Learning Approach to Forecasting Saltwater Intrusion in Chesapeake Bay and Its Major Tributaries.

Wegman, T. M., Pietrzak, J. D., Horner-Devine, A. R., Dijkstra, H. A., & Ralston, D. K. (2025). Observations of estuarine salt intrusion dynamics during a prolonged drought event in the Rhine-Meuse Delta. *Journal of Geophysical Research: Oceans*, 130(1), e2024JC021655. doi: 10.1029/2024JC021655

Werner, A. D., Bakker, M., Post, V. E., Vandenbohede, A., Lu, C., Ataie-Ashtiani, B., ... & Barry, D. A. (2013). Seawater intrusion processes, investigation and management: Recent advances and future challenges. *Advances in water resources*, 51, 3-26. doi: 10.1016/j.advwatres.2012.03.004

White, E., & Kaplan, D. (2017). Restore or retreat? Saltwater intrusion and water management in coastal wetlands. *Ecosystem Health and Sustainability*, 3(1), e01258. doi: 10.1002/ehs2.1258

Wong, K. C., & Valle-Levinson, A. (2002). On the relative importance of the remote and local wind effects on the subtidal exchange at the entrance to the Chesapeake Bay. *Journal of Marine Research*, 60(3), 477–498.

Yang, J., & Zhang, W. (2023). Storm-induced saltwater intrusion responds divergently to sea level rise in a complicated estuary. *Environmental Research Letters*, 19(1), 014011. doi: 10.1088/1748-9326/ad0e32

Yi, S., & Kondolf, G. M. (2024). Environmental planning and the evolution of inter-basin water transfers in the United States. *Frontiers in Environmental Science*, 12, 1489917. doi: 10.3389/fenvs.2024.1489917

Yokoi, S., Bressan, R.A., & Hasegawa, P.M. (2002). Salt stress tolerance of plants. *JIRCAS working report*, *23*(1), pp.25-33.

Zamrsky, D., Oude Essink, G. H., & Bierkens, M. F. (2024). Global impact of sea level rise on coastal fresh groundwater resources. *Earth's Future*, 12(1), e2023EF003581. doi: 10.1029/2023EF003581

**Supplementary**

Fig. S1. Observed annual salinity trends in the recent 40 years vs. rebuilt SWI length in the past century at the Delaware Bay

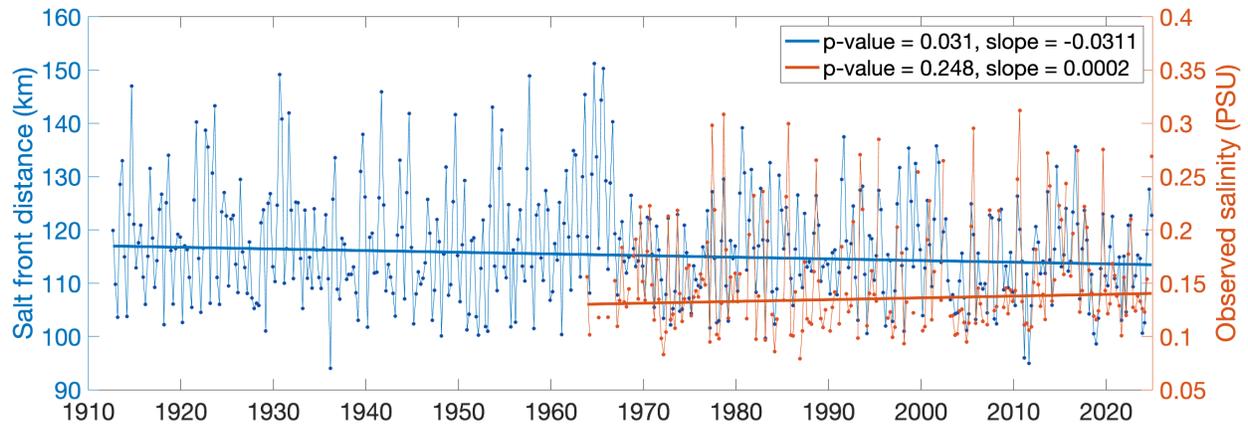

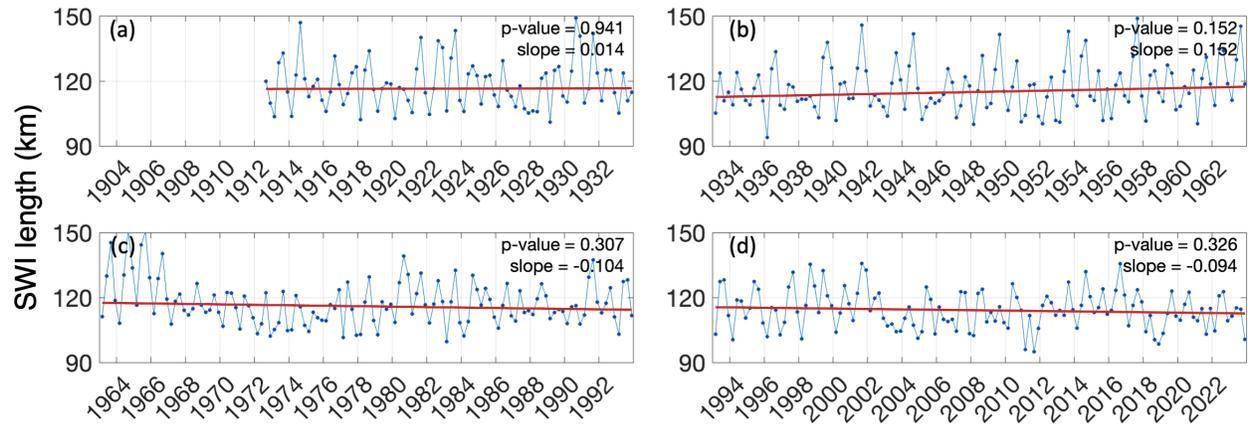

Fig. S2. Decadal trend slopes of reconstructed SWI length in Delaware Bay

Fig. S3. ML model training and tests

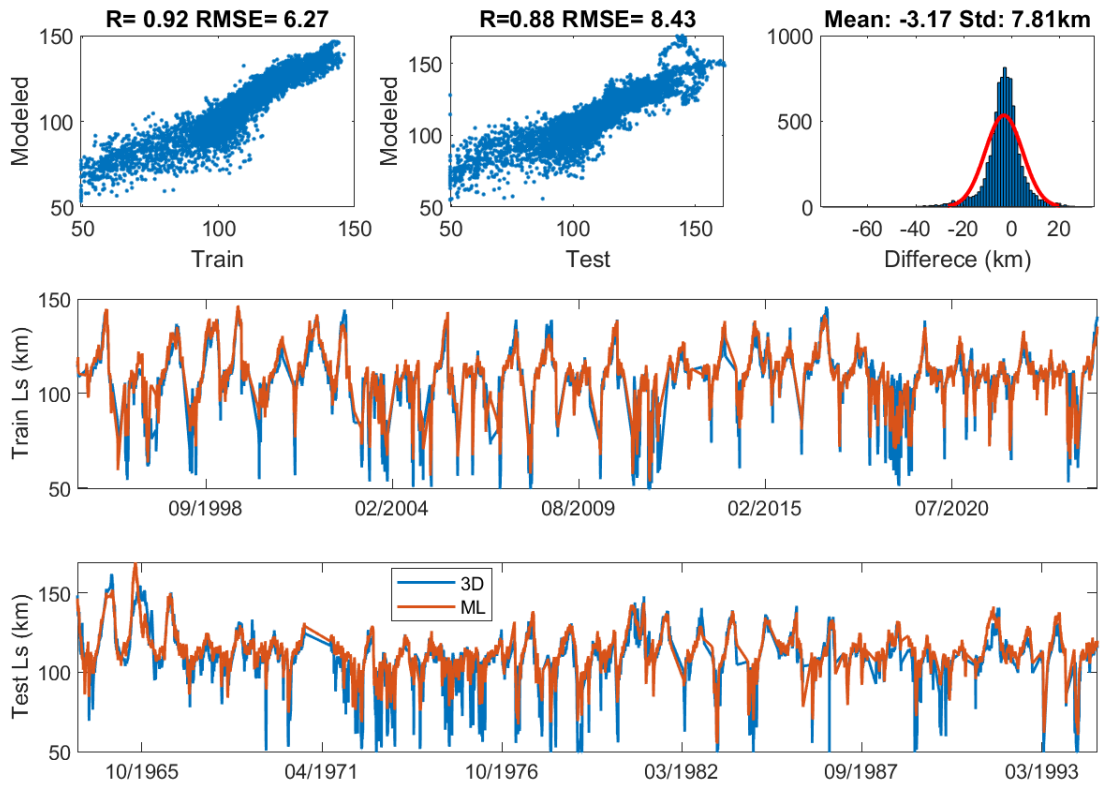

Table S1. Observed salinity trends at U.S. coastal stations (USGS and NERRS), based on available records beginning between the 1990s and 2010s through the present.

| Longitude | Latitude | Significance | Slope | Station Name |
|---|---|---|---|---|
| -72.841 | 41.45 | p-value > 0.25, unlikely | 0 | 01196500 |
| -73.955 | 41.386 | p-value > 0.25, unlikely | -0.001 | 01374019 |
| -74.778 | 40.222 | p-value < 0.05, significant | 0 | 01463500 |
| -75.139 | 39.946 | p-value < 0.05, significant | 0 | 01467200 |
| -75.203 | 39.879 | p-value > 0.25, unlikely | 0 | 01474703 |
| -75.366 | 39.837 | p-value < 0.05, significant | 0.001 | 01477050 |
| -75.568 | 39.501 | p-value > 0.25, unlikely | 0.005 | 01482800 |
| -76.886 | 40.255 | p-value > 0.25, unlikely | 0 | 01570500 |
| -77.128 | 38.95 | p-value > 0.25, unlikely | 0 | 01646500 |
| -77.529 | 38.308 | p-value < 0.05, significant | 0 | 01668000 |
| -77.332 | 37.768 | p-value < 0.05, significant | -0.001 | 01673000 |
| -78.086 | 37.671 | p-value > 0.25, unlikely | 0 | 02035000 |
| -78.086 | 37.671 | p-value > 0.25, unlikely | -0.001 | 02081094 |
| -79.044 | 33.833 | p-value < 0.05, significant | -0.001 | 02110704 |
| -78.753 | 33.799 | p-value > 0.25, unlikely | 0 | 02110755 |
| -78.867 | 33.741 | p-value < 0.05, significant | 0 | 02110760 |
| -78.719 | 33.821 | p-value > 0.25, unlikely | 0 | 02110770 |
| -78.656 | 33.852 | p-value < 0.25, possible | 0.021 | 02110777 |
| -79.174 | 33.445 | p-value < 0.05, significant | 0 | 02110815 |
| -79.949 | 33.094 | p-value < 0.05, significant | 0 | 02172020 |
| -79.958 | 33.057 | p-value < 0.05, significant | 0 | 02172040 |
| -79.936 | 33.058 | p-value > 0.25, unlikely | 0 | 02172050 |
| -79.923 | 32.984 | p-value < 0.05, significant | 0.029 | 02172053 |
| -81.151 | 32.235 | p-value < 0.05, significant | 0 | 02198840 |
| -81.155 | 32.165 | p-value < 0.05, significant | 0.035 | 02198920 |
| -81.683 | 29.596 | p-value < 0.05, significant | -0.002 | 02244040 |
| -81.342 | 25.829 | p-value > 0.25, unlikely | 0 | 02290930 |
| -91.192 | 30.446 | p-value < 0.25, possible | -0.001 | 07374000 |
| -89.564 | 29.634 | p-value < 0.05, significant | -0.238 | 07374526 |
| -89.606 | 29.586 | p-value < 0.05, significant | -0.285 | 07374527 |
| -90.921 | 29.249 | p-value < 0.25, possible | -0.054 | 07381349 |
| -91.88 | 29.713 | p-value < 0.25, possible | -0.026 | 07387040 |
| -92.136 | 29.674 | p-value > 0.25, unlikely | 0.003 | 07387050 |
| -93.247 | 30.237 | p-value < 0.25, possible | -0.034 | 08017044 |
| -93.3 | 30.032 | p-value < 0.25, possible | -0.074 | 08017095 |
| -93.349 | 29.816 | p-value < 0.25, possible | -0.086 | 08017118 |
| -122.127 | 38.045 | p-value < 0.05, significant | 0.273 | 11455780 |
| -122.226 | 38.061 | p-value < 0.05, significant | 0.227 | 11455820 |
| -72.553 | 41.542 | p-value < 0.25, possible | 0 | 01193050 |
| -72.346 | 41.312 | p-value > 0.25, unlikely | -0.014 | 01194796 |
| -73.71 | 40.866 | p-value < 0.05, significant | 0.065 | 01302250 |
| -73.593 | 40.905 | p-value < 0.25, possible | 0.029 | 01302845 |
| -73.143 | 40.963 | p-value < 0.05, significant | 0.053 | 01304057 |
| -72.307 | 41.137 | p-value < 0.05, significant | -0.058 | 01304200 |
| -72.639 | 40.918 | p-value > 0.25, unlikely | 0 | 01304562 |
| -74.28 | 40.992 | p-value < 0.25, possible | 0.001 | 01388000 |
| -74.122 | 40.147 | p-value < 0.05, significant | 0.001 | 01408029 |
| -75.801 | 39.962 | p-value < 0.05, significant | 0.002 | 01480617 |
| -75.673 | 39.969 | p-value < 0.05, significant | 0.002 | 01480870 |
| -75.593 | 39.87 | p-value < 0.05, significant | 0.002 | 01481000 |
| -75.577 | 39.77 | p-value < 0.05, significant | 0.002 | 01481500 |
| -75.614 | 39.466 | p-value > 0.25, unlikely | -0.002 | 01483177 |
| -75.458 | 39.011 | p-value > 0.25, unlikely | 0.009 | 01484080 |
| -75.291 | 38.595 | p-value < 0.05, significant | 0 | 01484525 |
| -75.099 | 38.625 | p-value > 0.25, unlikely | 0.021 | 01484680 |
| -77.58 | 36.331 | p-value < 0.05, significant | -0.001 | 0208062765 |
| -81.118 | 32.186 | p-value < 0.05, significant | -0.009 | 021989784 |
| -81.118 | 32.171 | p-value < 0.05, significant | -0.018 | 021989791 |
| -80.324 | 32.494 | p-value < 0.25, possible | 0.043 | acebbwq |
| -80.366 | 32.636 | p-value > 0.25, unlikely | -0.041 | acefcwq |
| -80.438 | 32.556 | p-value < 0.25, possible | -0.104 | acemcwq |
| -80.361 | 32.528 | p-value > 0.25, unlikely | 0.004 | acespwq |
| -84.88 | 29.702 | p-value > 0.25, unlikely | -0.028 | apacpwq |
| -85.058 | 29.675 | p-value > 0.25, unlikely | -0.013 | apadbwq |
| -84.875 | 29.786 | p-value > 0.25, unlikely | 0.023 | apaebwq |
| -84.875 | 29.786 | p-value > 0.25, unlikely | 0.025 | apaeswq |
| -76.721 | 38.796 | p-value < 0.05, significant | 0.001 | cbmipwq |
| -76.707 | 38.743 | p-value < 0.05, significant | 0 | cbmmcwq |
| -76.275 | 39.451 | p-value < 0.05, significant | 0.003 | cbmocwq |
| -76.714 | 38.781 | p-value < 0.05, significant | 0.002 | cbmrrwq |
| -76.611 | 37.347 | p-value > 0.25, unlikely | 0.034 | cbvcbwq |
| -76.393 | 37.216 | p-value < 0.25, possible | -0.026 | cbvgiwq |
| -76.714 | 37.415 | p-value > 0.25, unlikely | 0.027 | cbvtcwq |
| -75.636 | 39.389 | p-value < 0.25, possible | 0.009 | delblwq |
| -75.519 | 39.164 | p-value < 0.05, significant | 0 | deldswq |
| -75.499 | 39.114 | p-value < 0.25, possible | 0.035 | delllwq |
| -75.461 | 39.085 | p-value < 0.25, possible | 0.049 | delslwq |
| -121.754 | 36.846 | p-value > 0.25, unlikely | 0.011 | elkapwq |
| -121.738 | 36.835 | p-value > 0.25, unlikely | -0.026 | elknmwq |
| -121.739 | 36.818 | p-value < 0.05, significant | 0.06 | elksmwq |
| -121.779 | 36.811 | p-value < 0.25, possible | -0.013 | elkvmwq |
| -88.436 | 30.384 | p-value < 0.05, significant | -0.143 | gndbcwq |
| -88.405 | 30.418 | p-value < 0.05, significant | -0.143 | gndbhwq |
| -88.463 | 30.357 | p-value < 0.05, significant | -0.16 | gndblwq |
| -88.419 | 30.349 | p-value < 0.05, significant | -0.216 | gndpcwq |
| -70.869 | 43.072 | p-value < 0.05, significant | 0.08 | grbgbwq |
| -70.934 | 43.08 | p-value > 0.25, unlikely | 0.034 | grblrwq |
| -70.912 | 43.052 | p-value > 0.25, unlikely | 0.062 | grbsqwq |
| -81.246 | 29.737 | p-value < 0.05, significant | -0.074 | gtmfmwq |
| -81.257 | 29.667 | p-value < 0.25, possible | -0.115 | gtmpcwq |
| -81.367 | 30.051 | p-value < 0.05, significant | -0.329 | gtmpiwq |
| -81.307 | 29.869 | p-value < 0.05, significant | -0.06 | gtmsswq |
| -73.915 | 42.017 | p-value > 0.25, unlikely | 0 | hudskwq |
| -73.925 | 42.037 | p-value < 0.05, significant | 0 | hudtnwq |
| -73.926 | 42.027 | p-value < 0.25, possible | 0 | hudtswq |
| -74.339 | 39.508 | p-value < 0.25, possible | 0.011 | jacb6wq |
| -74.381 | 39.498 | p-value < 0.05, significant | 0.057 | jacb9wq |
| -74.552 | 39.594 | p-value < 0.05, significant | 0.025 | jacbawq |
| -74.461 | 39.548 | p-value < 0.05, significant | 0.053 | jacnewq |
| -66.239 | 17.943 | p-value < 0.25, possible | -0.022 | job09wq |
| -66.258 | 17.939 | p-value < 0.25, possible | 0.021 | job10wq |
| -66.229 | 17.943 | p-value > 0.25, unlikely | -0.001 | job19wq |
| -66.211 | 17.93 | p-value > 0.25, unlikely | 0.002 | job20wq |
| -151.409 | 59.602 | p-value < 0.25, possible | -0.007 | kachdwq |
| -151.721 | 59.441 | p-value > 0.25, unlikely | -0.006 | kacsdwq |
| -151.721 | 59.441 | p-value > 0.25, unlikely | 0.008 | kacsswq |
| -96.829 | 28.138 | p-value < 0.25, possible | -0.162 | marmbwq |
| -97.201 | 28.084 | p-value > 0.25, unlikely | 0.025 | marcwwq |
| -71.324 | 41.625 | p-value < 0.05, significant | 0.053 | narncwq |
| -71.341 | 41.641 | p-value > 0.25, unlikely | 0.009 | narpcwq |
| -71.321 | 41.578 | p-value > 0.25, unlikely | -0.003 | nartbwq |
| -71.321 | 41.578 | p-value > 0.25, unlikely | 0.005 | nartswq |
| -79.193 | 33.334 | p-value < 0.05, significant | -0.051 | niwcbwq |
| -79.167 | 33.36 | p-value < 0.25, possible | -0.028 | niwdcwq |
| -79.189 | 33.349 | p-value > 0.25, unlikely | 0.014 | niwolwq |
| -79.256 | 33.299 | p-value > 0.25, unlikely | 0.007 | niwtawq |
| -77.941 | 33.94 | p-value < 0.05, significant | -0.149 | nocecwq |
| -77.833 | 34.172 | p-value < 0.05, significant | 0.084 | noclcwq |
| -77.85 | 34.156 | p-value < 0.05, significant | 0.141 | nocrcwq |
| -77.935 | 33.955 | p-value < 0.25, possible | -0.07 | noczbwq |
| -82.512 | 41.349 | p-value < 0.05, significant | -0.002 | owcbrwq |
| -82.514 | 41.382 | p-value < 0.05, significant | -0.003 | owcolwq |
| -82.514 | 41.383 | p-value < 0.25, possible | -0.001 | owcwmwq |
| -122.531 | 48.556 | p-value > 0.25, unlikely | 0.003 | pdbbpwq |
| -122.502 | 48.496 | p-value < 0.05, significant | 0.011 | pdbbywq |
| -122.573 | 48.558 | p-value > 0.25, unlikely | -0.006 | pdbgswq |
| -81.477 | 25.892 | p-value > 0.25, unlikely | -0.054 | rkbfbwq |
| -81.516 | 25.901 | p-value < 0.25, possible | -0.119 | rkbfuwq |
| -81.734 | 26.027 | p-value < 0.05, significant | -0.122 | rkblhwq |
| -81.595 | 25.934 | p-value < 0.05, significant | -0.103 | rkbmbwq |
| -81.24 | 31.444 | p-value < 0.05, significant | -0.073 | sapcawq |
| -81.279 | 31.39 | p-value < 0.05, significant | -0.201 | sapdcwq |
| -81.273 | 31.479 | p-value < 0.05, significant | -0.092 | saphdwq |
| -81.296 | 31.418 | p-value < 0.05, significant | -0.067 | sapldwq |
| -122.033 | 38.195 | p-value < 0.25, possible | -0.049 | sfbfmwq |
| -122.014 | 38.184 | p-value < 0.05, significant | -0.077 | sfbsmwq |
| -122.509 | 38.016 | p-value < 0.25, possible | 0.11 | sfbgcwq |
| -122.46 | 38.001 | p-value > 0.25, unlikely | 0.037 | sfbccwq |
| -124.321 | 43.338 | p-value > 0.25, unlikely | -0.012 | soschwq |
| -124.322 | 43.317 | p-value > 0.25, unlikely | -0.011 | sosvawq |
| -124.32 | 43.282 | p-value < 0.05, significant | 0.166 | soswiwq |
| -117.129 | 32.559 | p-value < 0.05, significant | -0.03 | tjrbrwq |
| -117.131 | 32.568 | p-value > 0.25, unlikely | -0.008 | tjroswq |
| -117.116 | 32.601 | p-value > 0.25, unlikely | 0.019 | tjrsbwq |
| -70.587 | 43.298 | p-value < 0.05, significant | 0.016 | welhtwq |
| -70.563 | 43.32 | p-value < 0.05, significant | -0.05 | welinwq |
| -87.823 | 30.416 | p-value > 0.25, unlikely | -0.005 | wkbfrwq |
| -87.834 | 30.396 | p-value > 0.25, unlikely | -0.021 | wkbmbwq |
| -87.818 | 30.39 | p-value > 0.25, unlikely | -0.058 | wkbmrwq |
| -87.832 | 30.381 | p-value > 0.25, unlikely | -0.025 | wkbwbwq |
| -70.531 | 41.58 | p-value < 0.05, significant | 0.05 | wqbcrwq |
| -70.549 | 41.553 | p-value > 0.25, unlikely | 0.004 | wqbmhwq |
| -70.522 | 41.569 | p-value > 0.25, unlikely | 0.001 | wqbmpwq |